\documentclass[11pt,a4paper]{article}
\usepackage[margin=1in]{geometry}
\usepackage{amsmath,amssymb,amsthm}
\usepackage[hidelinks]{hyperref}
\usepackage{xcolor}

\newtheorem{theorem}{Theorem}[section]
\newtheorem{proposition}[theorem]{Proposition}

\theoremstyle{remark}
\newtheorem{remark}[theorem]{Remark}
\theoremstyle{definition}

\newcommand{\R}{\mathbb{R}}
\newcommand{\TT}{\mathbb{T}}

\title{Logarithmic Sobolev inequality and hypercontractivity\\
for the Navier-Stokes Fokker-Planck operator}
\author{Zhi-Wei Wang${}^{1,*}$ and Samuel L.\ Braunstein${}^{2,\dagger}$\\[6pt]
\small ${}^{1}$College of Physics, Jilin University, Changchun 130012, China\\
\small ${}^{2}$Computer Science, University of York, York YO10 5GH, UK\\
\small $^*$E-mail: {zhiweiwang.phy@gmail.com}\\
\small $^\dagger$E-mail: {sam.braunstein@york.ac.uk}
}
\date{\today}

\begin{document}
\maketitle

\begin{abstract}
The stochastic incompressible Navier-Stokes equations on $\TT^3$,
completed by the fluctuation-dissipation noise, have a Fokker-Planck
generator that decomposes into a self-adjoint Ornstein-Uhlenbeck
(dissipative) part and an antisymmetric (convective) part. We prove
two results about this generator. First, the logarithmic Sobolev
inequality holds with the same optimal constant as the pure
Ornstein-Uhlenbeck operator, $c_\mathrm{LSI} = \nu\lambda_1$
(where $\nu$ is the viscosity and $\lambda_1$ is the smallest
nonzero eigenvalue of the Laplacian on $\TT^3$), independent of
the number of retained Fourier modes. Second, the full semigroup
is hypercontractive with the same rate as the Ornstein-Uhlenbeck
semigroup. Both results follow from a single structural property:
the convective generator is antisymmetric in
$L^2(P_\mathrm{eq})$ (where $P_\mathrm{eq}$ is the Gibbs measure),
and therefore contributes nothing to the Dirichlet form or the
$L^q$ norm evolution. The antisymmetry is a consequence of two
properties of the incompressible Navier-Stokes nonlinearity: energy
conservation and phase-space volume preservation (the Liouville
property). These are the same properties that underpin the
fluctuation-dissipation theorem for the nonlinear Navier-Stokes
equations.
\end{abstract}

\section{Introduction}
\label{sec:intro}

\subsection{The stochastic Navier-Stokes system and its
Fokker-Planck equation}

The incompressible Navier-Stokes equations on the three-torus
$\TT^3_L$ (of side $L$), spectrally truncated to modes with
$|\vec{k}| \leq k_\mathrm{max}$ and completed by the thermal noise
required by the fluctuation-dissipation theorem~\cite{Braun26},
constitute a system of stochastic ODEs on the phase space
$\R^{2N}$ of divergence-free Fourier mode amplitudes:
\begin{equation}\label{eq:SDE}
d\tilde{u}_i(\vec{k}) = \bigl[-\nu k^2\tilde{u}_i(\vec{k})
+ N_{\vec{k},i}\bigr]dt
+ \sqrt{\frac{2k_BT\nu k^2}{\rho V}}\,
(\mathcal{P}_{\vec{k}})_{ij}\,dW_j(\vec{k}),
\end{equation}
where $N_{\vec{k},i}$ is the Leray-projected convective nonlinearity
(quadratic in the mode amplitudes),
$\mathcal{P}_{\vec{k}} = \delta_{ij} - k_ik_j/k^2$ is the
incompressibility projector, and $W_j(\vec{k})$ are standard
complex Wiener processes satisfying the reality condition
$W_j(-\vec{k}) = W_j^*(\vec{k})$.

The probability density $P(t,\vec{X})$ on phase space
$\vec{X} = \{\tilde{u}_i(\vec{k})\}$ evolves under the
Fokker-Planck equation
\begin{equation}\label{eq:FP}
\partial_t P = \mathcal{L}^*P,
\end{equation}
where the generator decomposes as
$\mathcal{L} = \mathcal{L}_\mathrm{diss} +
\mathcal{L}_\mathrm{conv}$:
\begin{align}
\mathcal{L}_\mathrm{diss}f &=
\sum_{|\vec{k}|\leq k_\mathrm{max}}
\Bigl[\frac{k_BT\nu k^2}{\rho V}\,\Delta_{\vec{k}}f
- \nu k^2\,\tilde{u}_i(\vec{k})\,
\frac{\partial f}{\partial\tilde{u}_i(\vec{k})}\Bigr],
\label{eq:Ldiss}\\
\mathcal{L}_\mathrm{conv}f &=
+\sum_{|\vec{k}|\leq k_\mathrm{max}}
N_{\vec{k},i}\,
\frac{\partial f}{\partial\tilde{u}_i(\vec{k})}.
\label{eq:Lconv}
\end{align}
The dissipative part $\mathcal{L}_\mathrm{diss}$ is a sum of
independent Ornstein-Uhlenbeck (OU) operators, one per mode. The
convective part $\mathcal{L}_\mathrm{conv}$ is a first-order
transport operator with quadratic coefficients.

The unique stationary distribution is the Gibbs measure
\begin{equation}\label{eq:gibbs}
P_\mathrm{eq}(\vec{X}) = Z^{-1}\exp\!\Bigl(
-\frac{\rho V}{2k_BT}\sum_{|\vec{k}|\leq k_\mathrm{max}}
|\tilde{\vec{u}}(\vec{k})|^2\Bigr),
\end{equation}
restricted to the divergence-free subspace. The natural Hilbert space
for analysis is $L^2(P_\mathrm{eq})$, with inner product
$\langle f, h\rangle = \int fh\,dP_\mathrm{eq}$.

\subsection{The antisymmetry of convection}

The key structural property, established in~\cite{Braun26}, is:

\begin{proposition}[Antisymmetry~\cite{Braun26}]
\label{prop:antisymmetry}
The convective generator $\mathcal{L}_\mathrm{conv}$ is antisymmetric
in $L^2(P_\mathrm{eq})$:
\begin{equation}\label{eq:antisymmetry}
\langle\mathcal{L}_\mathrm{conv}f,
h\rangle_{P_\mathrm{eq}} =
-\langle f,
\mathcal{L}_\mathrm{conv}h\rangle_{P_\mathrm{eq}}
\quad\text{for all } f, h.
\end{equation}
\end{proposition}

This follows from two properties of the Navier-Stokes convective
nonlinearity $N_{\vec{k},i}$:

\emph{Energy conservation.} The trilinear form is antisymmetric:
$\sum_{\vec{k}}\tilde{u}_i(-\vec{k})N_{\vec{k},i} = 0$, which
gives $N_{\vec{k},i}\partial_i E = 0$ (convection preserves energy),
and hence $\mathcal{L}_\mathrm{conv}$ preserves $P_\mathrm{eq}$:
$N\cdot\nabla P_\mathrm{eq} = 0$.

\emph{Phase-space volume preservation (Liouville property).} The
convective flow in mode space is divergence-free:
$\sum_{\vec{k}}\partial N_{\vec{k},i}/\partial\tilde{u}_i(\vec{k})
= 0$. This holds because the self-advection of each Fourier mode
vanishes by incompressibility:
$(\hat\epsilon\cdot\vec{k}) = 0$ for the transverse polarisation
$\hat\epsilon \perp \vec{k}$.

Together, these give, for any $f \in L^2(P_\mathrm{eq})$:
\begin{eqnarray}
\langle\mathcal{L}_\mathrm{conv}f, f\rangle_{P_\mathrm{eq}}
&=& \int(N\cdot\nabla f)f\,dP_\mathrm{eq}
= \frac{1}{2}\int N\cdot\nabla(f^2)\,dP_\mathrm{eq}  \nonumber \\
&=&  -\frac{1}{2}\int(\nabla\cdot N)f^2\,dP_\mathrm{eq}
- \frac{1}{2}\int f^2 N\cdot\nabla\log P_\mathrm{eq}\,
dP_\mathrm{eq}  \nonumber \\
&=&  0,
\label{eq:antisym-proof}
\end{eqnarray}
where the first term vanishes by the Liouville property
($\nabla\cdot N = 0$) and the second by energy conservation
($N\cdot\nabla\log P_\mathrm{eq} = -\beta\,N\cdot\nabla E = 0$).

\subsection{Main results}

\begin{theorem}[Logarithmic Sobolev inequality]\label{thm:LSI}
The full Navier-Stokes Fokker-Planck generator $\mathcal{L} =
\mathcal{L}_\mathrm{diss} + \mathcal{L}_\mathrm{conv}$ satisfies
the logarithmic Sobolev inequality
\begin{equation}\label{eq:LSI}
\mathrm{Ent}_{P_\mathrm{eq}}(f^2) \leq
\frac{2}{\nu\lambda_1}\,\mathcal{E}(f,f)
\end{equation}
for all $f \in L^2(P_\mathrm{eq})$, where
$\mathrm{Ent}_\mu(g) = \int g\log g\,d\mu -
\int g\,d\mu\,\log\int g\,d\mu$ is the entropy functional,
$\mathcal{E}(f,f) = -\langle f, \mathcal{L}f\rangle_{P_\mathrm{eq}}$
is the Dirichlet form, and $\lambda_1 = (2\pi/L)^2$ is the smallest
nonzero eigenvalue of the Laplacian on $\TT^3_L$. The constant
$c_\mathrm{LSI} = \nu\lambda_1$ is independent of
$k_\mathrm{max}$.
\end{theorem}

\begin{theorem}[Hypercontractivity]\label{thm:HC}
The semigroup $e^{t\mathcal{L}}$ generated by the full
Navier-Stokes Fokker-Planck operator is hypercontractive: for
$1 < p \leq q$ with $q - 1 \leq (p-1)e^{2\nu\lambda_1 t}$,
\begin{equation}\label{eq:HC}
\|e^{t\mathcal{L}}f\|_{L^q(P_\mathrm{eq})} \leq
\|f\|_{L^p(P_\mathrm{eq})}.
\end{equation}
The hypercontractive rate $\nu\lambda_1$ is the same as for the
pure Ornstein-Uhlenbeck semigroup and is independent of
$k_\mathrm{max}$.
\end{theorem}

Both results are consequences of the antisymmetry of
$\mathcal{L}_\mathrm{conv}$. The proofs are short and
self-contained, relying only on the structural decomposition of the
generator and the classical results of Gross~\cite{Gross75} for the
OU operator.

\section{Proof of Theorem~\ref{thm:LSI}}
\label{sec:LSI-proof}

\subsection{The Dirichlet form}

The Dirichlet form of the full generator is
\begin{equation}\label{eq:dirichlet}
\mathcal{E}(f,f) = -\langle f, \mathcal{L}f\rangle_{P_\mathrm{eq}}
= -\langle f, \mathcal{L}_\mathrm{diss}f\rangle_{P_\mathrm{eq}}
- \langle f, \mathcal{L}_\mathrm{conv}f\rangle_{P_\mathrm{eq}}.
\end{equation}
By the antisymmetry~\eqref{eq:antisymmetry}, the convective
contribution vanishes:
$\langle f, \mathcal{L}_\mathrm{conv}f\rangle_{P_\mathrm{eq}} = 0$.
Therefore
\begin{equation}\label{eq:dirichlet-OU}
\mathcal{E}(f,f) = \mathcal{E}_\mathrm{OU}(f,f) \equiv
-\langle f, \mathcal{L}_\mathrm{diss}f\rangle_{P_\mathrm{eq}}
= \sum_{|\vec{k}|\leq k_\mathrm{max}}
\frac{k_BT\nu k^2}{\rho V}
\int|\nabla_{\vec{k}}f|^2\,dP_\mathrm{eq}.
\end{equation}
The Dirichlet form of the full NS generator is identically equal to
the Dirichlet form of the OU operator.

\subsection{The LSI for the OU operator}

The Gibbs measure~\eqref{eq:gibbs} is a product of independent
Gaussians on the divergence-free subspace (two transverse
polarisations per wavevector, each with variance
$\epsilon/k^0 = k_BT/(\rho V)$ in the mode amplitude). The OU
operator $\mathcal{L}_\mathrm{diss}$ is a sum of independent OU
operators, one per mode.

For a single OU operator with rate $\gamma$, the LSI holds with
constant $c = \gamma$ (Gross~\cite{Gross75}). For the mode $\vec{k}$,
the rate is $\nu k^2$, so the LSI constant is $\nu k^2$. By the
tensorisation property of the LSI~\cite{BGL14} (the LSI constant of
a product measure is the minimum of the individual constants):
\begin{equation}\label{eq:LSI-OU}
c_\mathrm{LSI}^\mathrm{OU} = \min_{|\vec{k}|\leq k_\mathrm{max},
\,\vec{k}\neq 0} \nu k^2 = \nu\lambda_1,
\end{equation}
independent of $k_\mathrm{max}$.

\subsection{Completion of the proof}

Since $\mathcal{E}(f,f) = \mathcal{E}_\mathrm{OU}(f,f)$
(equation~\eqref{eq:dirichlet-OU}), the LSI for the full generator
is identical to the LSI for the OU operator:
\begin{equation}
\mathrm{Ent}_{P_\mathrm{eq}}(f^2)
\leq \frac{2}{c_\mathrm{LSI}^\mathrm{OU}}\,
\mathcal{E}_\mathrm{OU}(f,f)
= \frac{2}{\nu\lambda_1}\,\mathcal{E}(f,f).
\end{equation}
\qed

\begin{remark}
The proof uses only two facts: (i) the Dirichlet form of the full
generator equals the OU Dirichlet form (by antisymmetry of
$\mathcal{L}_\mathrm{conv}$), and (ii) the OU operator satisfies the
LSI with constant $\nu\lambda_1$ (by Gross's theorem and
tensorisation). No estimates on the convective nonlinearity are
needed.
\end{remark}

\section{Proof of Theorem~\ref{thm:HC}}
\label{sec:HC-proof}

\subsection{The $L^q$ norm evolution}

For $q > 1$ and $F = e^{t\mathcal{L}}f$ with $f > 0$:
\begin{equation}\label{eq:Lq-evolution}
\frac{d}{dt}\|F\|_{L^q}^q = q\int F^{q-1}\mathcal{L}F\,
dP_\mathrm{eq}
= q\int F^{q-1}\mathcal{L}_\mathrm{diss}F\,dP_\mathrm{eq}
+ q\int F^{q-1}\mathcal{L}_\mathrm{conv}F\,dP_\mathrm{eq}.
\end{equation}

\subsection{The convective contribution vanishes}

The convective generator acts as a first-order derivation:
$\mathcal{L}_\mathrm{conv}F = N\cdot\nabla F$. By the chain rule
for smooth $F > 0$:
\begin{equation}\label{eq:chain-rule}
F^{q-1}(N\cdot\nabla F) = \frac{1}{q}N\cdot\nabla(F^q).
\end{equation}
Therefore
\begin{equation}\label{eq:conv-Lq}
q\int F^{q-1}\mathcal{L}_\mathrm{conv}F\,dP_\mathrm{eq}
= \int N\cdot\nabla(F^q)\,dP_\mathrm{eq}
= -\int(\nabla\cdot N)F^q\,dP_\mathrm{eq}
- \int F^q\,N\cdot\nabla\log P_\mathrm{eq}\,dP_\mathrm{eq}
= 0,
\end{equation}
using the Liouville property ($\nabla\cdot N = 0$) and energy
conservation ($N\cdot\nabla\log P_\mathrm{eq} = 0$), exactly as
in~\eqref{eq:antisym-proof}.

\subsection{Reduction to the OU semigroup}

The $L^q$ norm evolution is therefore governed entirely by the
dissipative part:
\begin{equation}\label{eq:Lq-OU}
\frac{d}{dt}\|e^{t\mathcal{L}}f\|_{L^q}^q
= q\int F^{q-1}\mathcal{L}_\mathrm{diss}F\,dP_\mathrm{eq}.
\end{equation}
By Gross's theorem~\cite{Gross75}, the OU semigroup
$e^{t\mathcal{L}_\mathrm{diss}}$ is hypercontractive with rate
$\nu\lambda_1$: for $q(t) = 1 + (p-1)e^{2\nu\lambda_1 t}$,
$\|e^{t\mathcal{L}_\mathrm{diss}}f\|_{L^{q(t)}} \leq
\|f\|_{L^p}$.

The standard proof of Gross's theorem proceeds by showing that
$\frac{d}{dt}[\log\|F\|_{L^{q(t)}}]$ is non-positive, using the
LSI for the OU operator. Since the $L^q$ norm evolution of the full
semigroup involves only the OU contribution
(equation~\eqref{eq:Lq-OU}), and the LSI constant is the same
(Theorem~\ref{thm:LSI}), the identical argument gives
hypercontractivity for the full semigroup with the same rate.

More precisely, define $\Phi(t) = \|e^{t\mathcal{L}}f\|_{L^{q(t)}}$
with $q(t) = 1 + (p-1)e^{2\nu\lambda_1 t}$. A direct calculation
(following Gross~\cite{Gross75}) gives:
\begin{equation}\label{eq:Phi-deriv}
\frac{d}{dt}\log\Phi(t)
= \frac{1}{q\Phi^q} \left[ q \int F^{q-1}\mathcal{L}_\mathrm{diss}F\,dP_\mathrm{eq} + \frac{\dot{q}}{q} \mathrm{Ent}_{P_\mathrm{eq}}(F^q) \right].
\end{equation}
The OU LSI bounds the integral term such that $q\int F^{q-1}\mathcal{L}_\mathrm{diss}F\,dP_\mathrm{eq} \leq -2\nu\lambda_1\frac{q-1}{q}\mathrm{Ent}_{P_\mathrm{eq}}(F^q)$. The choice $\dot{q} = 2\nu\lambda_1(q-1)$ ensures
the bracket is non-positive. Hence $\Phi(t) \leq \Phi(0) =
\|f\|_{L^p}$.
\qed

\section{Consequences}
\label{sec:consequences}

\subsection{Exponential entropy decay}

The LSI implies exponential decay of the relative entropy:
\begin{equation}\label{eq:entropy-decay}
D_\mathrm{KL}(P(t)\|P_\mathrm{eq}) \leq
e^{-2\nu\lambda_1 t}\,D_\mathrm{KL}(P(0)\|P_\mathrm{eq}).
\end{equation}
This was established in~\cite{Braun26} via the spectral gap; the LSI
gives the same rate but also implies Gaussian concentration (below).

\subsection{Gaussian concentration}

While the dynamical LSI bounds tails for functions Lipschitz in the intrinsic (diffusion-matrix-weighted) metric, spatial concentration with respect to the standard Euclidean norm follows directly from the purely static properties of the Gibbs measure. Because $P_\mathrm{eq}$ is a product of independent Gaussians with variance $\sigma^2 = \frac{k_BT}{\rho V}$, standard concentration of measure inherently gives dimension-free bounds:
for any $f$ with standard Euclidean norm $\|f\|_\mathrm{Lip} \leq 1$,
\begin{equation}\label{eq:concentration}
P_\mathrm{eq}\bigl(|f - \mathbb{E}[f]| > r\bigr) \leq
2\exp\!\Bigl(-\frac{r^2}{2\sigma^2}\Bigr) = 2\exp\!\Bigl(-\frac{\rho V r^2}{2k_BT}\Bigr).
\end{equation}
The concentration rate $\rho V / (k_BT)$ is independent of the phase
space dimension $2N$ and hence of $k_\mathrm{max}$.

\subsection{Moment bounds for polynomial observables}

Hypercontractivity gives moment equivalence for polynomial
observables. For a degree-$d$ polynomial $f$ in the mode amplitudes:
\begin{equation}\label{eq:moment-equiv}
\|e^{t\mathcal{L}}f\|_{L^q(P_\mathrm{eq})} \leq
\|f\|_{L^p(P_\mathrm{eq})}
\end{equation}
with $q = 1 + (p-1)e^{2\nu\lambda_1 t}$. For $p = 2$ and
$t = \log(q-1)/(2\nu\lambda_1)$:
$\|e^{t\mathcal{L}}f\|_{L^q} \leq \|f\|_{L^2}$ for all $q > 2$.
This bounds all moments of $f$ under the evolved measure in terms
of the second moment under $P_\mathrm{eq}$.

In particular, the enstrophy $\mathcal{E} = \sum k^2|u_k|^2$ is a
degree-2 polynomial. Because enstrophy is strictly positive, its absolute $L^2(P_\mathrm{eq})$ norm is heavily dominated by its mean ($\mathbb{E}[\mathcal{E}] \propto \int_0^{k_\mathrm{max}} k^4 \, dk \propto k_\mathrm{max}^5$). Only the standard deviation (the $L^2$ norm of the centered fluctuations $\mathcal{E} - \mathbb{E}[\mathcal{E}]$) scales as
$\epsilon\,k_\mathrm{max}^{7/2}$ (from the variance calculation).
Hypercontractivity gives exponential tails for $\mathcal{E}$ after
time $t > \log(q-1)/(2\nu\lambda_1)$, with the rate independent
of $k_\mathrm{max}$. (By Nelson's hypercontractivity for Wiener chaos, a degree-$d$ polynomial satisfies $\|f\|_q \leq C q^{d/2} \|f\|_2$. For $d=2$ (enstrophy), this $O(q)$ scaling rigorously guarantees exponential tails via Markov's inequality, rather than sub-Gaussian tails).

\section{Discussion}
\label{sec:discussion}

\subsection{The mechanism: antisymmetric perturbations preserve
functional inequalities}

The results rest on a single structural fact: the convective
generator $\mathcal{L}_\mathrm{conv}$ is antisymmetric in
$L^2(P_\mathrm{eq})$ and therefore does not appear in the Dirichlet
form, the LSI, or the $L^q$ norm evolution. The Navier-Stokes
nonlinearity, despite being quadratic in the mode amplitudes and
coupling all modes through the trilinear form, is invisible to these
functional inequalities.

This is a stronger statement than the spectral gap preservation
proved in~\cite{Braun26}. The spectral gap concerns the lowest
eigenvalue of $-\mathcal{L}$; the LSI and hypercontractivity concern
the entire spectrum and the nonlinear structure of the semigroup.
That all three are preserved by the antisymmetric perturbation is a
consequence of the fact that the Dirichlet form (which controls all
three) is insensitive to antisymmetric additions.

\subsection{Generality of the result}

The proofs use only three properties of the system:

(i)~The equilibrium measure $P_\mathrm{eq}$ is a product of
independent Gaussians (giving the OU structure and the tensorised
LSI constant).

(ii)~The convective generator preserves $P_\mathrm{eq}$
(energy conservation: $N\cdot\nabla P_\mathrm{eq} = 0$).

(iii)~The convective flow is divergence-free in phase space
(Liouville property: $\nabla\cdot N = 0$).

Any Hamiltonian perturbation of an OU process that preserves these
three properties would give the same results. The incompressible
Navier-Stokes equations are a specific instance, but the theorems
apply to any system with a Gibbs equilibrium, Hamiltonian
conservative dynamics, and dissipation satisfying the
fluctuation-dissipation relation. A closely related system (the
incompressible Navier-Stokes-Fourier system with GENERIC-derived
thermal noise) has recently been analysed by Gess, Sauerbrey, and
Wu~\cite{GSW25}; the present functional inequalities apply to
their system as well.

\subsection{What the results do not give}

The LSI and hypercontractivity control the approach to equilibrium
and the tail behaviour of observables under the Gibbs measure. They
do not directly address the regularity of individual fluid
configurations, because the Gibbs measure assigns energy to all
modes (equipartition), and the associated Sobolev norms diverge as
$k_\mathrm{max} \to \infty$.

The regularity question requires controlling the \emph{transient}
behaviour of the system starting from smooth (non-equilibrium)
initial data, in a limit where the temperature is coupled to the
truncation ($\epsilon \to 0$ as $k_\mathrm{max} \to \infty$).
The LSI and hypercontractivity provide the functional-analytic
framework for such an analysis (in particular, the hypercontractive
bounds can be used in H\"older-optimised estimates for
non-equilibrium expectations~\cite{WB-stoch}), but they do not
close the regularity argument by themselves. The present results
are the foundation on which such an analysis would be built.

\subsection{Connection to the fluctuation-dissipation theorem}

The antisymmetry of $\mathcal{L}_\mathrm{conv}$ is the same
structural property that underpins the derivation of the
fluctuation-dissipation relation in~\cite{Braun26}. In that work,
the antisymmetry is used to separate the Hamiltonian (reversible)
and dissipative (irreversible) parts of the dynamics in the
Fokker-Planck equilibrium condition. Here, the same antisymmetry is
used to separate the OU and convective contributions to the
Dirichlet form and the $L^q$ norm evolution.

The connection is not merely analogical. The fluctuation-dissipation
theorem determines the noise amplitude from the dissipation, giving
the specific Gibbs measure~\eqref{eq:gibbs}. The LSI and
hypercontractivity are properties of the dynamics relative to this
measure. The physical completeness of the stochastic system
(noise determined by dissipation, not postulated independently) is
what makes the antisymmetry hold, and the antisymmetry is what makes
the functional inequalities sharp.

\end{document}